\documentclass[preprint2]{emulateapj}

\slugcomment{version 2010 Jan 25: fm}

\shorttitle{Gamma Ray Burst in the \fer~ era}
\shortauthors{F. Massaro, J. E. Grindlay, A. Paggi, 2009}

\newcommand{\fer}{{\it Fermi}}

\begin{document}

\title{Gamma Ray Bursts in the \fer~ era: \\ the spectral energy distribution of the prompt emission}%@title

\author{F. Massaro\altaffilmark{1}, J. E. Grindlay\altaffilmark{1}, A. Paggi\altaffilmark{2}}

\affil{Harvard - Smithsonian Astrophysical Observatory, 60 Garden Street, Cambridge, MA 02138}
\affil{Dipartimento di Fisica, Universit\`{a} di Roma Tor Vergata, Via della Ricerca scientifica 1, I-00133 Roma, Italy}

\begin{abstract} %@abs
Gamma Ray Bursts (GRBs) show evidence of different light curves, duration, afterglows, 
host galaxies and they explode within a wide redshift range. 
However, their spectral energy distributions (SEDs) appear to be very similar showing a curved shape.
Band et al. (1993) proposed a phenomenological description of the integrated spectral shape 
for the GRB prompt emission, the so called {\it Band function}.
In this letter we suggest an alternative scenario to explain the curved shape of GRB SEDs: the {\it log-parabolic model}.
In comparison with the Band spectral shape our model is statistically favored because it fits the GRB spectra
with one parameter less than the Band function and it is motivated by a theoretical acceleration scenario.
The new \fer~ observations of GRBs will be crucial to disentangle between these two models.
\end{abstract}


\section{Introduction}
The physical mechanisms behind the GRBs prompt emission are still under debate.
Band et al. (1993), investigating the BATSE GRBs sample proposed a 
phenomenological description of the integrated spectral shape 
for the GRB prompt emission, the so called {\it Band function}.
The introduction of this function was strongly suggested by the observational evidence that the 
shape of the Spectral energy Distribution (SED) of the GRB prompt emission is convex and broadly peaked.
It is remarkable that there has not been physical explanation in terms of 
accelerations processes and non-thermal radiative losses
that can led to the Band spectral shape.

In this letter we propose to describe the curved shape of GRB prompt 
emission using the {\it log-parabolic} function (Massaro et al. 2004),
successfully used to describe the SEDs of BL Lac objects over several decades.
First, we consider the differences between this model in comparison 
with the Band function investigating the different $\gamma$-ray flux predictions
in the \fer~LAT energy range.
Second, we point toward the physical interpretation of the log-parabolic shape in terms of Fermi acceleration mechanisms.
Finally, we present a simple synchrotron emission model to explain the GRB prompt emission,
that appears to be the most reasonable scenario.

For our numerical results, we use cgs units unless stated otherwise
and we assume a flat cosmology with $H_0=72$ km s$^{-1}$ Mpc$^{-1}$,
$\Omega_{M}=0.27$ and $\Omega_{\Lambda}=0.73$ (Spergel et al. 2007).

\section{The shape of the Spectral Energy Distribution}
GRBs have a non-thermal spectrum that varies strongly from one burst to another. 
It is generally found that a simple power law does not fit well their spectra because 
of a steepening toward the high energies.
The Band function phenomenological model (Band et al. 1993) 
describes the prompt time-integrated GRBs spectra, composed 
by two power laws joined smoothly at a break energy $E_b$:
\begin{equation}
F(E)=\left\{
\begin{array} {lllllll}
F_0\left(\frac{E}{E_0}\right)^{\alpha}~exp\left(-\frac{E}{E_c}\right)&&& && E\leq E_b\\
F_1\left(\frac{E}{E_0}\right)^{\beta}& &&&&E\geq E_b\\ 
\end{array}
\right .
\end{equation}
where $F(E)$ is the number of photons per unit of area and energy and time, 
while $E_0$ is a reference energy usually fixed to the value of 100 keV. 
Under the continuity requirement for the function $F(E)$ and its first derivative, 
the break energy and normalization are given by:
\begin{eqnarray}
E_b&=&(\alpha-\beta)~E_c \\
F_1&=&F_0~\left[\frac{(\alpha-\beta)E_c}{E_0}\right]^{\alpha-\beta}~e^{(\beta-\alpha)}
\end{eqnarray}

There are no particular theoretical scenarios that predict this spectral
shape making it only a phenomenological model.
However, it provides good fits to most of the observed spectra
in terms of four parameters, namely: the two photon indices $\alpha$ and
$\beta$, the exponential cut-off $E_c$ and the normalization constant
$F_0$, with all four parameters directly estimated during the fitting procedure. 
The peak energy $E_p$ of the SED (i.e. S(E)=E$^2$F(E)) is related to the
spectral parameters by:
\begin{equation}
E_p=(\alpha+2)E_c < E_c
\end{equation}
for which, typically, $-2<\alpha<-1$. 
We also note that for typical values $\alpha\simeq -1.4$ and $\beta\simeq -2.4$ (Band et al. 1993), 
$E_{b}\approx E_{c}$.

We propose to describe and interpret the shape of the 
SED in the GRB prompt emission using a model defined by the following equation:
\begin{equation}
F (E) = F_0 \left(\frac{E}{E_0}\right)^{-a-b~log\left(E/E_0\right)}
\end{equation}
where F$_0$ is the normalization, $a$ is the spectral index at energy $E_0$
and $b$ is the parameter which measures the spectral curvature.
This model is known as {\it log-parabolic}, a curve having a parabolic shape in a log-log plot 
(Massaro et al. 2004, Massaro et al. 2006).
We remark that this spectral distribution is the classical log-normal statistical distribution. 
%-----------------------------------------------------------------------------------------------------------------------
\begin{figure}[!htp]
\includegraphics[height=7.5cm,width=8.5cm,angle=0]{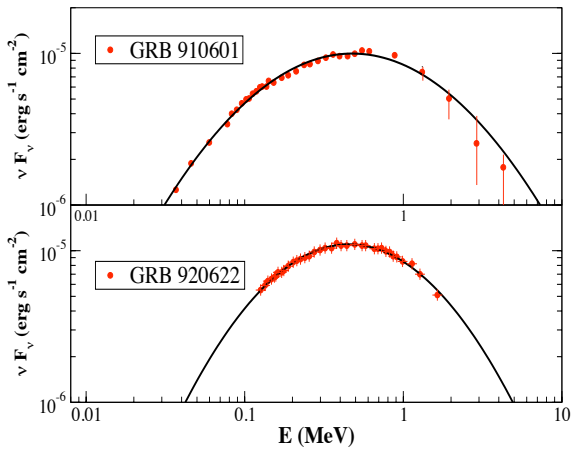}
\caption{Two SEDs of GRB 910601 (upper panel) and 
GRB 920622 (lower panel) with the log-parabolic best fit model (black line).
It is clear how the spectral shape is well described by 
the log-parabolic function (data taken from Schaefer  et al. 1994 and from Tavani et al. 1996).}
\end{figure}
%-----------------------------------------------------------------------------------------------------------------------
In particular, for this function, it is possible to define an energy dependent photon index $\Gamma (E)$
given by the log-derivative of Eq. (5),
\begin{equation}
\Gamma(E) = a + 2~b~log(E/E_0)
\end{equation}
which describes the continuous change in the spectral slope.

The peak energy $E_p$ and the height of the SED S(E) calculated at its peak frequency $S_p$
can be evaluated by the following relations
\begin{eqnarray}
E_p&=&E_0 10^{\frac{2-a}{2b}}\\
S_p&=&S_0 10^{\frac{(2-a)^2}{4b}}
\end{eqnarray}
where $S_0$ is $S(E_0)$. 

Consequently the spectral shape can be expressed in terms of 
$b$, $E_p$ and $S_p$ using the relation:
\begin{equation}
S(E) = E^2 F(E) = S_p~10^{-b~\log^2(E/E_{p})}~, 
\end{equation}
where  $S_p=E_{p}^{2}\, F(E_p)$. In this form the values of the parameters $b$, $E_p$ and 
$S_p$ are estimated independently in the fitting procedure, whereas those derived from Eq. (7) and Eq. (8) are
affected by intrinsic correlations (Tanihata et al. 2004, Tramacere et al. 2007).

As example of the goodness of the fitting procedure, we show in Fig. 1 the best fits of two GRB SEDs evaluated with the log-parabolic model
namely GRB 910601 and GRB 920622, two of the most bright and well studied GRBs present in literature,
(data taken from Schaefer et al. 1994 and Tavani et al. 1996). 
Their best fit parameters are: E$_p = 0.47 \pm 0.01$ MeV, $b = 0.74 \pm 0.03$, 
S$_p = 9.99 \pm 0.12 \times 10^{-6}$ erg s$^{-1}$ cm$^{-2}$ for GRB 910601
while E$_p = 0.460 \pm 0.003$ MeV, $b = 0.95 \pm 0.02$, 
S$_p = 1.098 \pm 0.005 \times 10^{-5}$ erg s$^{-1}$ cm$^{-2}$  for GRB 920622.

We note that GRB spectra appear to be narrower with respect to the BL Lac objects (Massaro et al. 2008),
having the curvature parameters close to 1.

\section{Log-parabolic {\it vs} Band model}
From a statistical point of view, the log-parabolic shape requires one parameter less than the Band function
and so is favored.
There are two main differences between these two models:
first, the slope at low energies of the Band function is a power-law while the log-parabolic one
has a milder curvature and second, the high energy tail is naturally curved and the expected flux in the $\gamma$-ray band
is lower than the one predicted by the Band spectral shape (see Fig. 2). 
In particular, the log-parabolic model can describe a continuous curvature over the whole spectrum while the Band function
can only mimic it around the SED peak. 
%-----------------------------------------------------------------------------------------------------------------------
\begin{figure}[!htp]
\includegraphics[height=6.5cm,width=8.5cm,angle=0]{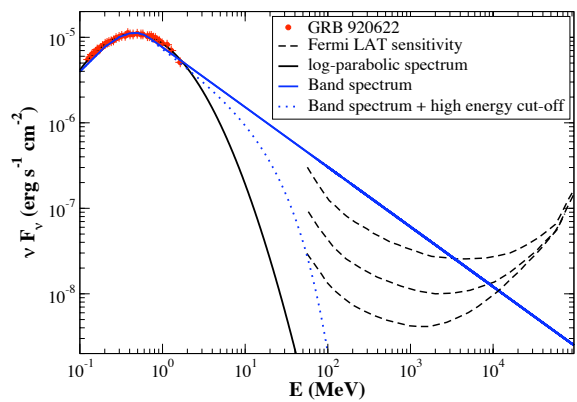}
\caption{The SED of GRB 920622 and the comparison between the Band and the log-parabolic function 
used to describe its emission. The extrapolation of the Band function predicts that a similar GRBs should be detected within 100s
in the \fer~LAT energy range, while the log-parabolic model is more in agreement with the GRB detection rate
of the first year of \fer~LAT observations. The \fer~LAT sensitivity has been evaluated from that reported in Atwood et al. (2009).}
\end{figure}
%-----------------------------------------------------------------------------------------------------------------------

In the recent \fer~observations only 9 GRBs have been detected at high energies in the LAT energy range (Granot et al. 2009)
in comparison with the predictions provided by the extrapolation of the Band function (Omodei et al. 2007, Band et al. 2009, Omodei et al. 2009). 
Several explanations have been proposed to correct the expectations and a high energy cut-off 
has been introduced in the Band function to arrange the lack of the observed GRBs in the \fer~LAT band (Band et al. 2009).
The introduction of this exponential cut-off increases the number of parameters in the Band function while
the log-parabolic model appears to have a natural explanation for the \fer~observations without the introduction of any
new spectral parameter. 

In Fig. 2 we plot the SED for the GRB 920622 with different spectral model extrapolations in the \fer~LAT energy range.
The \fer~LAT sensitivity evaluated for an exposure of 100 s and with three different backgrounds as reported
in Atwood et al. (2009) is also shown.
Using the Band model a typical GRB with a similar spectral shape and total fluence of GRB 920622, 
as those reported in the BATSE bright GRB catalog (Schaefer et al. 1994),
is expected to be detected in the \fer~LAT energy range while the predictions of the log-parabolic shape are very different
and no LAT detection is expected for this GRB.

We fitted with the log-parabolic model the SEDs of all brightest GRBs present in the BATSE catalog (Schaefer et al. 1994) 
detected during 1991. The shape of the SED is well described in terms of this model as already shown in Fig. 1.
We found that extrapolating the log-parabolic spectrum only 2 GRBs, out of 20 total, are expected to be 
marginally detected in the \fer~LAT band within 100s, 
more in agreement with the detection rate of the first year \fer~observations
with respect to the high \fer~LAT detection rate estimated using the Band model extrapolation (Omodei et al. 2009, Band et al. 2009).
%-----------------------------------------------------------------------------------------------------------------------
\begin{figure}[!htp]
\includegraphics[height=6.5cm,width=8.5cm,angle=0]{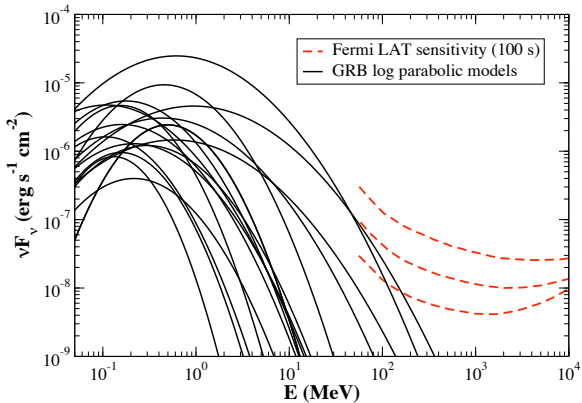}
\caption{The log-parabolic models of the 20 brightest GRBs in the BATSE catalog detected during 1991 (Schaefer et al. 1994)
It is clear how the extrapolation of the log-parabolic function led to conclude that only few GRBs are expected to have a detection in the \fer~LAT
energy range in contrast with the expectations of the Band model.}
\end{figure}
%-----------------------------------------------------------------------------------------------------------------------
In Fig. 3, It is evident how the curved shape described by a log-parabolic function is more in agreement with the \fer~LAT
first year GRB detections.
As in Fig. 2, the \fer~LAT sensitivity has been evaluated from that reported in Atwood et al. (2009) rescaled for 100s.

\section{Acceleration mechanisms and synchrotron radiation in the prompt emission}
The energy spectrum of accelerated particles by some statistical mechanism, 
such as those occurring in shock waves, is usually written as a power law. 
The origin of this interpretation resides in the first order Fermi 
acceleration mechanism (Bell 1978, Blandford \& Eichler 1987 and Protheroe 2004), 
originally presented to explain the cosmic ray spectrum.
However, the observational evidence that the SED of BL Lacs objects has a curved shape  
has demanded toward a different interpretation.
In particular, Landau et al. (1986) provided a useful description of the synchrotron component of 
BL Lac objects in terms of a log-parabolic model, that has been recently applied to describe 
the synchrotron X-ray spectra of TeV BL Lacs (Massaro et al. 2008).

The theoretical interpretation of the log-parabolic model resides in the energy distribution of the emitting particles.
The general solution of the kinetic equation, for the time-dependent distributions with respect to the energy,
yields curved particle energy distributions (PEDs) in the form of log-normal function,
where terms taking into account of the stochastic and systematic acceleration by the Fermi mechanisms
are considered (Kardashev 1962).

Assuming a simple $\delta$ function as initial condition for the PED,
the analytical solution of the particle kinetic equation yields the log-parabolic shape in the form:
\begin{equation}\label{logpar}
N\left({\gamma}\right)= N_0\, {\left({\frac{\gamma}{\gamma_0}}\right)}^{-s-r\log{\left(\gamma/\gamma_0\right)}}
\end{equation}
where the parameters $s$, $r$ and the normalization $N_0$ are directly linked to the physical parameters $\lambda_1$ and $\lambda_2$, 
while $\gamma_0$ is a reference energy.
We note that the PED curvature $r$ is only directly linked to the diffusion coefficient,
this means that the curved shape of the particle distribution depends on considering   
the second order Fermi acceleration mechanisms (Kardashev 1962, Paggi et al. 2009).
The case $r=0$ corresponds to the simple power-law electron distribution as expected 
by a first order acceleration mechanism.
The mean quadratic energy of the PED, $\langle{\gamma^2}\rangle$, corresponds to the second normalized momentum
 and can be expressed in the form:
\begin{equation}
\langle{\gamma^2}\rangle = \gamma_0 10^{(2-s)/2r} = \gamma_p
\end{equation} 
which corresponds to the peak of $\gamma^2~N(\gamma)$.
By freezing the PED slope, $s$, to the value of 2, as expected in the first order Fermi acceleration mechanism,
it is possible to describe its shape in terms of the PED curvature $r$, $\gamma_p$ and the normalization $N_0$
and Eq. (8) can be written in the form
\begin{equation}\label{logpar}
N\left({\gamma}\right)= N_0\, {\left({\frac{\gamma}{\gamma_p}}\right)}^{-2-r\log{\left(\gamma/\gamma_p\right)}}.
\end{equation}

Finally, we remark that it has been recently shown that including synchrotron and inverse Compton radiative losses as well as 
of the ``disappearance" of fast particles that escape from the acceleration region, 
either as a result of nuclear collisions or escape from the acceleration region,
the numerical solution for the kinetic equation can be successfully described in terms of a log-parabolic shape 
(Tramacere et al. 2009, Paggi et al. 2009).

Under this assumption that the synchrotron radiation is emitted by a log-parabolic 
PED (Eq. 10), the resulting flux density and consequently 
the SED can be well approximated by a log-parabolic shape, expressed as: Eq. (5).
%-----------------------------------------------------------------------------------------------------------------------
\begin{figure}[!htp]
\includegraphics[height=6.5cm,width=8.5cm,angle=0]{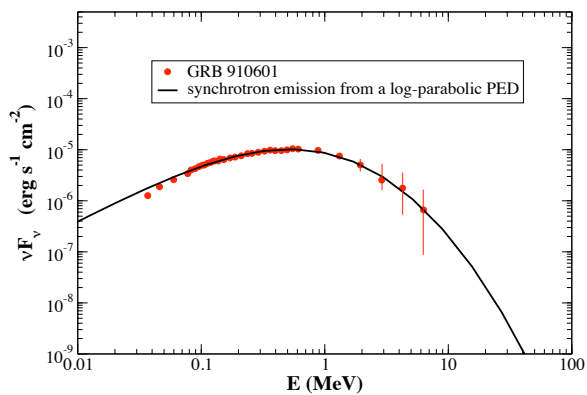}
\caption{The synchrotron model applied to the case of GRB 910601. 
The intrinsic source parameters are reported in Table 1.}
\end{figure}
%-----------------------------------------------------------------------------------------------------------------------
An alternative scenario is based on the assumption that the probabilty to accelerate particles depends on energy in 
a simple relation given by ${\rm P} \propto \gamma^{-q}$,
and in this case the resulting PED yields toward a log-parabolic shape (Massaro et al. 2004, 2006). 

Applying the numerical code developed by Massaro et al. (2007) and presented in Paggi et al. (2009), 
we calculated the synchrotron emission by a log-parabolic PED to describe the observed SED of GRB 910601.
This is a clear example of how the log-parabolic scenario successfully describes the shape of the GRB prompt emission.
The parameters assumed for our calculations are given in Tab. 1, and in  Fig. 4 we show 
the GRB 910601 SED with the model adopted.
We fixed the redshift of this GRB to 1 because it is unknown.
The parameters in Tab. 1 are all consistent with plausible values of the GRBs emitting region (e.g. M\'esz\'aros 2002 and zhand \& M\'esz\'aros 2002).
%-----------------------------------------------------------------------------------------------------------------------
\begin{table}
\caption{The intrinsic source parameters for GRB 910601.}
\begin{tabular}{lllc}
\hline
Parameter  & Symbol & units & value \\
\hline
\noalign{\smallskip}
redshift                       & z & --- & 1.0 \\
PED slope                 & s & --- & 2  \\
PED curvature          & r & --- & 7.58 \\
PED energy peak     & $\gamma_p$ & --- & 1.97$\times$ 10$^4$ \\
PED minimum energy     & $\gamma_{min}$ & --- & 10$^3$ \\
PED maximum energy    & $\gamma_{max}$ & --- & 5$\times$ 10$^8$ \\
electron density        & n$_{el}$ & cm$^{-3}$ & 1.13$\times$ 10$^4$ \\
beaming~factor        & $\delta$ & --- & 30 \\
magnetic~field          & B & Gauss & $10^4$ \\
volume                        & V & cm$^3$ & 10$^{42}$ \\
\noalign{\smallskip}
\hline
\end{tabular}\\
\end{table}
%-----------------------------------------------------------------------------------------------------------------------
The synchrotron model, evaluated with a log-parabolic PED, is in agreement with the data, 
appearing to be a good description of the GRB SEDs of the prompt emission.

\section{Conclusions}
In the present letter we propose to interpret the SEDs of GRBs prompt emission using the log-parabolic shape.
In comparison with the Band function (Band et al. 1993)
the log-parabolic shape is favored for two main reasons.

First, it is statistically better, because it requires only 3 parameters, namely 
the curvature $b$, the peak energy $E_p$ and the height of the SED evaluated at the peak energy $S_p$ 
(see Eq. 6 and Eq. 15), while the usual Band model needs 4 spectral parameters, namely: $\alpha$, $\beta$, $E_c$ and $F_0$
or 5 if another high energy exponential cut off is introduced.
Second, the proposed function has a strong physical motivation.
This shape is directly related to the solution of kinetic equation
for the particles accelerated by Fermi mechanisms when
the random acceleration is also taken into account with all the 
other terms (e.g. first order Fermi mechanisms) (Kardashev 1962).

In the recent \fer~LAT observations only few GRBs have been detected 
in contrast with the predictions of the Band function. 
The high energy curvature of the log-parabolic shape has a natural explanation for the \fer~observations 
without the introduction of any new parameter, as the exponential cut-off in the Band function.

As shown in Massaro et al. (2006) or more recenly in Tramacere et al. (2009) and Paggi et al. (2009),
the synchrotron emission of a log-parabolic electron distribution 
yields a curved SED near its peak, well described in terms of the same spectral shape.
In this letter, we also presented how the synchrotron scenario from a log-parabolic PED
can describe the spectrum of the GRBs prompt emission, successfully. 

Finally, we remark that a different scenario, including other synchrotron or inverse Compton components
and their spectral evolution with time, 
can make the spectral shape of the GRB prompt emission more complex in the \fer~LAT energy range
as for example recently observed in the case of GRB 090902B (Abdo et al. 2009).

\begin{acknowledgements}
F. Massaro acknowledges the Foundation BLANCEFLOR Boncompagni-Ludovisi, n'ee Bildt 
for the grant awarded him in 2009 to support his research. 
The work at SAO is supported by NASA-GRANT NNX10AD50G.
\end{acknowledgements}

{}

\end{document}